\documentclass{article}

    \PassOptionsToPackage{numbers, compress}{natbib}

\usepackage[table,xcdraw]{xcolor}

\usepackage[final]{neurips_2021}

\usepackage[utf8]{inputenc} 
\usepackage[T1]{fontenc}    
\usepackage{hyperref}       
\usepackage{url}            
\usepackage{booktabs}       
\usepackage{amsfonts}       
\usepackage{nicefrac}       
\usepackage{microtype}      
\usepackage{xcolor}         
\usepackage{graphicx}

\title{ Localized Perturbations For Weakly-Supervised Segmentation of Glioma Brain Tumours}

\usepackage{authblk}
\author[1,3]{Sajith M. Rajapaksa}
\author[1,2,3,4,5,6]{Farzad Khalvati}
\affil[1]{Neurosciences and Mental Health, The Hospital for Sick Children, Toronto, Canada}
\affil[2]{Department of Diagnostic Imaging, The Hospital for Sick Children}
\affil[3]{Institute of Medical Science, University of Toronto}
\affil[4]{Department of Medical Imaging, University of Toronto}
\affil[5]{Department of Mechanical and Industrial Engineering, University of Toronto}
\affil[6]{Vector Institute, Toronto, Canada}
\affil[ ]{\textit {sajith.rajapaksa@utoronto.ca, farzad.khalvati@utoronto.ca}}

\begin{document}
\bibliographystyle{IEEEtran.bst}

\maketitle

\begin{abstract}
Deep convolutional neural networks (CNNs) have become an essential tool in the medical imaging-based computer-aided diagnostic pipeline. However, training accurate and reliable CNNs requires large fine-grain annotated datasets. To alleviate this, weakly-supervised methods can be used to obtain local information from global labels. This work proposes the use of localized perturbations as a weakly-supervised solution to extract segmentation masks of brain tumours from a pretrained 3D classification model. Furthermore, we propose a novel optimal perturbation method that exploits 3D superpixels to find the most relevant area for a given classification using a U-net architecture. Our method achieved a Dice similarity coefficient (DSC) of 0.44 when compared with expert annotations. When compared against Grad-CAM, our method outperformed both in visualization and localization ability of the tumour region, with Grad-CAM only achieving 0.11 average DSC.

\end{abstract}

\appendix

\section{Introduction}

With the increased usage of medical imaging such as Magnetic Resonance Imaging (MRI) in diagnostic procedures of different diseases, to alleviate pressure from radiologists, demand for deep learning-based computer-aided diagnosis systems has also increased. These systems can fast-track examinations of potential malignant cases to provide patients with better care. However, training accurate and reliable convolutional neural networks (CNNs) requires large fine-grain annotated datasets (e.g., manual tumour annotation). Nevertheless, such datasets are not widely available mainly because manual annotation is prohibitively expensive. This opens the opportunity to explore weakly-supervised solutions where weak labels such as global classification can obtain fine-grain information like region of interest (ROI). Methods such as attention networks \cite{ding2019hierarchical} have been proposed as potential solutions. However, such solutions require modification to the classification model's architecture and retraining. Another method, Grad-CAM \cite{selvaraju2016grad}, struggles with 3D CNN architectures as it produces low-resolution outputs. This work proposes the classification relevance map, which is a model-agnostic weakly-supervised segmentation method that generates the ROI for an input image based on a novel optimal perturbation on superpixels. Furthermore, we show that this method can also be used as an effective visualization tool for 3D CNN architectures to improve interpretability for clinical usage by generating detailed ROI. In the following, we present a brief literature review of three main concepts of our work.

\textbf{Brain Tumour Classification} has been a key interest in the field of AI for medical imaging. Accurate and fast classification of tumours allows for selecting better care for the patient \cite{doolittle2004state}. As one of the most common brain tumours, Gliomas are split into two main subcategories depending on the severity, Low-Grade Glioma (LGG) and High-Grade Glioma (HGG). With the introduction of the Multimodal Brain Tumor Segmentation Challenge (BraTs) \cite{menze2014multimodal,bakas2017advancing,bakas2018identifying,bakas2017segmentation}, numerous successful methods have been proposed for classification \cite{amin2020brain,usman2017brain}. However, most of them rely on segmentation of the ROI before classification, which requires fine-grain expert annotations. Some 3D CNN-based approaches have been proposed to classify tumours using the whole volume directly. However, they fail to achieve a performance comparable to that of models trained on the ROIs \cite{rehman2021microscopic, ye2017glioma}. In this work, we apply our proposed method to generate ROIs from a CNN trained only on 3D volumes and global classification labels. 

\textbf{Superpixels} can be defined as groupings of perceptually similar pixels to create a meaningful image with fewer primitive elements for processing. The term coined by Ren and Malik in Learning a Classification Model for Segmentation set out to solve the problem that as pixels are not natural entities, they are meaningless as representations of images \cite{ren2003learning}. Superpixels have been used to segment medical images as they capture similar subregions in images such as tumours \cite{ouyang2020self,huang2020segmentation}. 

\textbf{Perturbation} in deep learning has been commonly used for model attacks and robustness evaluation. Local Interpretable Model-Agnostic Explanations (LIME) technique was proposed to use localized perturbations as a model agnostic method to increase interpretability by evaluating the model's confidence when images are perturbed \cite{ribeiro2016should}. Randomized Input Sampling for Explanation (RISE) further improved this work by introducing randomized global masking to calculate region relevancy to a classification \cite{petsiuk2018rise}. Both of these works showed high adaptability as these methods are model agnostic and can be easily used on any pretrained model. 

\section{Methods}

The Multimodal Brain Tumor Segmentation Challenge 2020 dataset was used as our primary dataset \cite{menze2014multimodal,bakas2017advancing,bakas2018identifying,bakas2017segmentation}. 3D volumes of T1w, T1wCE, T2w and FLAIR sequences were available for each patient along with the glioma tumour classification and segmentation. Only 190 3D scans (training : 133, validation : 19, test : 38 ) were used  to maintain a 60\% HGG and 40\% LGG tumour ratio. Each sequence was independently normalized using min-max normalization and center cropped to 128 $\times$ 128 $\times$ 128 volumes.

\begin{figure}[h]
\centering
\includegraphics[width=11cm]{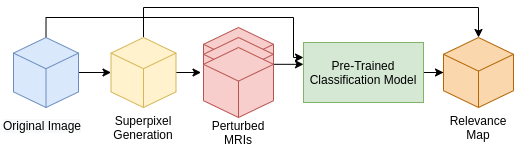}
\caption{Proposed relevance map generation pipeline.}
\end{figure}

As shown in Figure 1, our method is comprised of three main components. First, we generate a superpixel map for a given input. Secondly, we iterate through the superpixels to create a perturbed input to the trained CNN model. Finally, we measure the difference in the trained CNN probability for each perturbed input compared to that of the original image to generate the relevance map. We have selected a 3D Resnet \cite{he2015deep} model as our baseline classification architecture to evaluate the proposed method. The model was trained on all four sequences with a shape of 4 x 128 x 128 x 128 as input with a learning rate of 0.01 and the Adam optimizer \cite{kingma2014adam} for 100 epochs using binary cross-entropy as the loss function. The trained model at epoch 81 was selected for evaluation as it achieved the highest validation accuracy. The model achieved an area under the ROC curve of 0.83 on the test set. 

For 3D superpixel generation, Simple Linear Iterative Clustering (SLIC) was used \cite{achanta2012slic,van2014scikit}. We conducted a grid search as part of the experimentation to determine the best parameters for sequence type for superpixel generation and the number of superpixels. The best set of parameters were selected based on the Dice similarity coefficient achieved on tumour segmentation with optimal thresholding, where we assumed the best thresholding would be known. For a given superpixel, we perturbed the selected region on all four modalities to be used as input. We experimented using three trivial perturbation methods as baselines: Blank perturbation, where we set all pixels in the selected region to 0 similar to LIME, max and min perturbations by setting the pixel values in the selected region to the sequence's max and min value, respectively. 

To determine the optimal perturbation for a given segmented 3D superpixel, we propose a 3D U-net model \cite{ronneberger2015u} with a segmented superpixel as input and a perturbation mask as the output. The model was trained on superpixels generated with four randomly selected samples from the training set. Adam optimizer \cite{kingma2014adam} with perturbation loss defined as  $\sum \frac{1}{abs( non perturbed - perturbed)}$ was used for training, where the training is optimized to increase the difference between classification confidence of perturbed and non-perturbed input images.

Finally, relevance maps were generated by assigning each superpixel a score on its ability to change the model's confidence on classification by calculating the absolute difference between the perturbed and non perturbed images. Relevance maps were then normalized between 0 and 100, where regions with 0 score had no effect on the classification, and those with 100 had the highest impact. 

To evaluate generated relevance maps' accuracy, we first computed the average Dice similarity coefficient (DSC) with optimal thresholding, which clusters multiple superpixels. Secondly, we computed the average DSC for the highest-ranked superpixels. Finally, we compare our method against the visualization and segmentation that can be produced using Grad-CAM method. 

\section{Results}
Our grid search on superpixel parameters showed that 100 superpixels generated on the T2w sequence yield the best results. We also determined that the overall blank perturbation has the most significant effect on trivial perturbation methods with a DSC of 0.40 with optimal thresholding on the relevance map. Using the same superpixel parameters, the relevance map generated using our proposed optimal perturbation method achieved a DSC of 0.44. Table 1 shows the DSC for ranked superpixels where the blank perturbed superpixels generated on the T2w sequence achieved an average DSC of 0.25. The highest-ranking optimally perturbed superpixel achieved an average DCS of 0.31.

When compared against Grad-CAM, our method outperformed both in visualization quality and the ability to localize the tumour. With optimal thresholding, Grad-CAM achieved an average DSC 0f 0.11 while and our method achieved 0.44. Figure 2. shows a sample visualization where our relevance map generated a higher resolution and more meaningful visualization of the ROI.

\begin{table}[h!]
\centering
\caption{Average DSC  on ranked superpixels generated on different optimal and blank perturbation.}
\begin{tabular}{ccc}
\hline
Rank & \begin{tabular}[c]{@{}c@{}}Avg. DSC with \\ Optimal Perturbation\end{tabular} & \begin{tabular}[c]{@{}c@{}}Avg. DSC with \\ Blank Perturbation\end{tabular} \\ \hline
1    & 0.31                                                                          & 0.25                                                                        \\ \hline
2    & 0.18                                                                          & 0.07                                                                        \\ \hline
3    & 0.10                                                                          & 0.06                                                                        \\ \hline
\end{tabular}
\end{table}



\begin{figure}[h!]
\centering
\includegraphics[width=10cm]{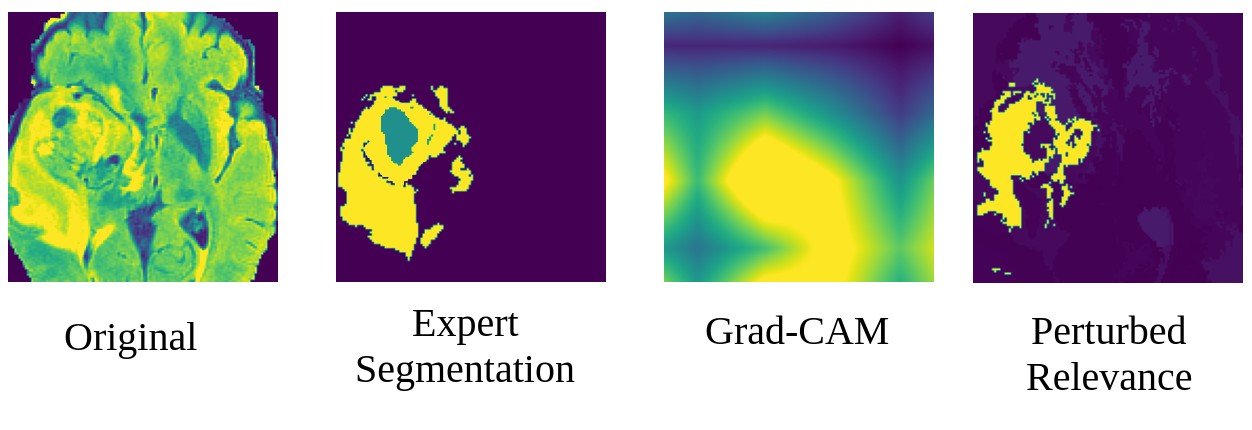}
\caption{Generated visualization comparison between expert annotated, Grad-CAM generated and our proposed method (Relevance Map).}
\end{figure}

\section{Conclusions}
In this work, we successfully implemented a novel localized perturbation method to extract the ROI (tumours) from a 3D CNN trained only for the classification of MRI brain glioma tumours. We also showed that our proposed method of superpixels-based perturbation masks generator (relevance map) could also generate visualization maps to significantly improve the interpretability of black-box classification models. For future work, we aim to improve the perturbation model by also incorporating global level perturbation. 

\section{Negative Societal Impact Statement }
Improving the interpretability of black-box models will allow attackers also to improve their methods. However, we believe the benefit of understanding these models considerably outweighs the perturbed attack concerns.

\begin{ack}
This research has been supported by Huawei Technologies Canada Co., Ltd.
\end{ack}

\bibliography{cite.bib}
\clearpage
\appendix
\section{Appendix}
\subsection{Relevance Maps for Understating Training Process}

\begin{figure}[h]
\centering
\includegraphics[width=14cm]{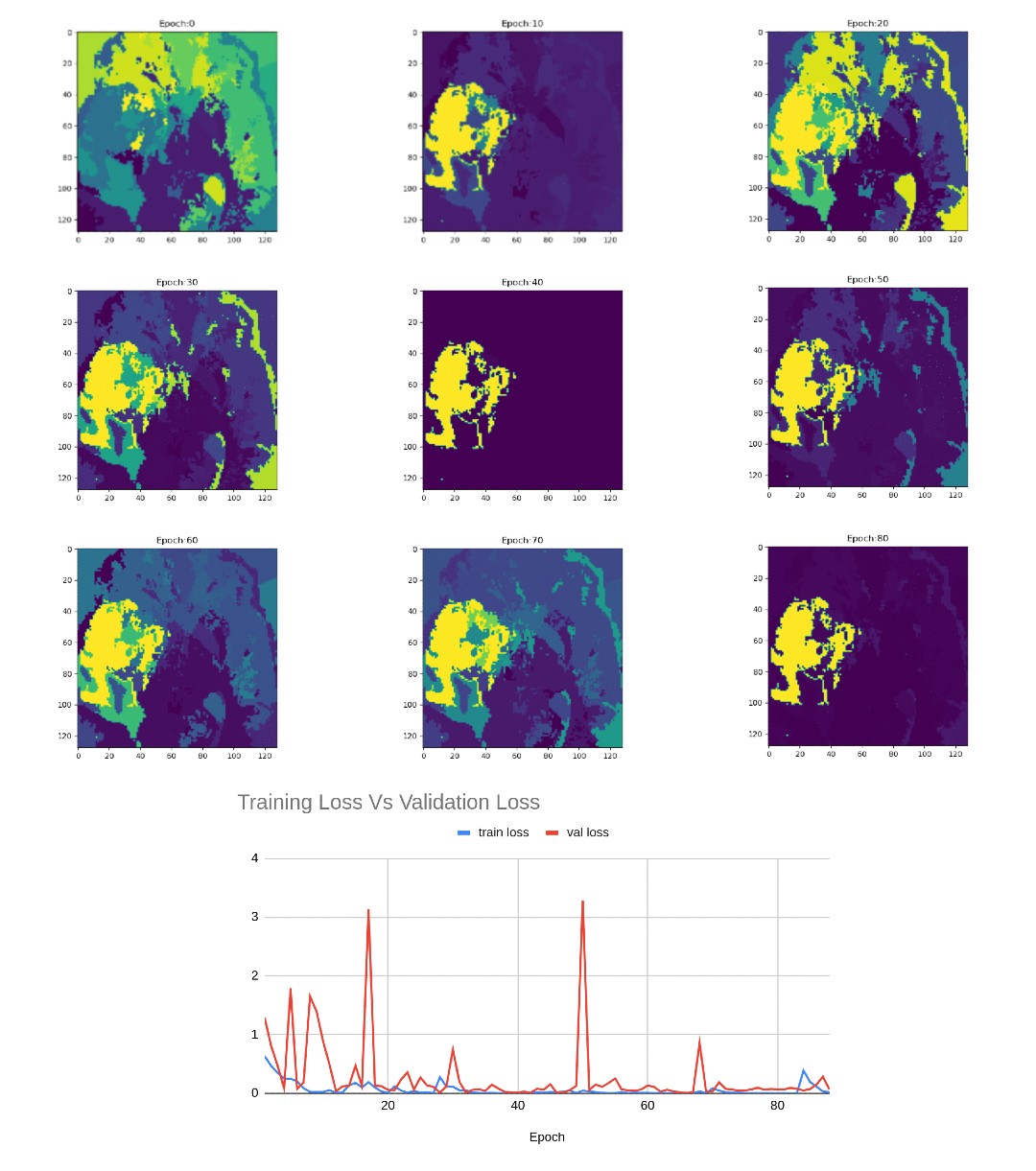}
\caption{Relevance maps for different epochs of training visualising different levels of focus.}
\end{figure}

Figure 3 displays the ability of our proposed method to highlight regions of interest while the training process is underway. It shows that as the validation loss decreases, the model focuses more on a single region of interest. When validation loss spikes, the model puts the relevance on a larger area. 

\clearpage
\subsection{ Dice Similarity Coefficient on Combined Superpixels}

\begin{table}[h]
\centering
\caption{Average DSC evaluation on cumulative top ranking superpixels}
\begin{tabular}{ccc}
\hline
Rank   & \begin{tabular}[c]{@{}c@{}}Avg. DSC with \\ Optimal Perturbation\end{tabular} & \begin{tabular}[c]{@{}c@{}}Avg. DSC with \\ Blank Perturbation\end{tabular} \\ \hline
1      & \cellcolor[HTML]{FFFFFF}0.31                                                  & \cellcolor[HTML]{FFFFFF}0.25                                                \\ \hline
1+2    & \cellcolor[HTML]{FFFFFF}\textbf{0.34}                                         & \cellcolor[HTML]{FFFFFF}0.22                                                \\ \hline
1+..+3 & \cellcolor[HTML]{FFFFFF}0.32                                                  & \cellcolor[HTML]{FFFFFF}0.18                                                \\ \hline
1+..+4 & \cellcolor[HTML]{FFFFFF}0.29                                                  & \cellcolor[HTML]{FFFFFF}0.16                                                \\ \hline
1+..+5 & \cellcolor[HTML]{FFFFFF}0.25                                                  & \cellcolor[HTML]{FFFFFF}0.16                                                \\ \hline
\end{tabular}
\end{table}

We compare the average DSC that can be achieved by combining superpixels ranked on relevance. As shown in Table A.2, our proposed method outperformed blank perturbation when combining the first two superpixels generating an average DSC of 0.34.



\subsection{Grid Search for Optimal Superpixel Parameters}
\begin{table}[h]
\centering
\begin{tabular}{ l r r r r r r r r r }
\hline
      & \multicolumn{3}{c}{Blank}                                                    & \multicolumn{3}{c}{Min}                                                      & \multicolumn{3}{c}{Max}                                                      \\ \hline
      & \multicolumn{1}{c }{50} & \multicolumn{1}{c}{100} & \multicolumn{1}{c}{250} & \multicolumn{1}{c }{50} & \multicolumn{1}{c}{100} & \multicolumn{1}{c}{250} & \multicolumn{1}{c }{50} & \multicolumn{1}{c}{100} & \multicolumn{1}{c}{250} \\ \hline
T1w    & 0.24                    & 0.27                     & 0.29                     & 0.13                    & 0.15                     & 0.17                     & 0.09                    & 0.1                      & 0.1                      \\ \hline
T1wCE  & 0.27                    & 0.27                     & 0.29                     & 0.17                    & 0.19                     & 0.21                     & 0.1                     & 0.1                      & 0.1                      \\ \hline
T2w    & \textbf{0.37}           & \textbf{0.4}             & \textbf{0.34}            & 0.21                    & 0.22                     & 0.24                     & 0.1                     & 0.1                      & 0.1                      \\ \hline
FLAIR & 0.33                    & 0.33                     & 0.32                     & 0.24                    & 0.22                     & 0.22                     & 0.1                     & 0.1                      & 0.1                      \\ \hline
\end{tabular}
\caption{Parameter grid search results.}
\end{table}

Table A3 shows DSCs obtained using optimal thresholding on different perturbation and superpixel generation parameters. We found Blank perturbation (among trivial methods) with 100 superpixels generated on the T2w sequence performed the best.
\end{document}